\documentclass[reprint]{JASA}
\graphicspath{{./fig/}}
\begin{document}

\title{Second register production on the clarinet: nonlinear losses in the register hole as the decisive physical phenomenon
}
\author{Nathan Szwarcberg}
\email{nathan.szwarcberg@buffetcrampon.com}
\affiliation{Buffet Crampon, 5 Rue Maurice Berteaux, 78711 Mantes-la-Ville, France}
\affiliation{Aix Marseille Univ, CNRS, Centrale Marseille, LMA, Marseille, France}
\author{Tom Colinot}
\affiliation{Buffet Crampon, 5 Rue Maurice Berteaux, 78711 Mantes-la-Ville, France}
\author{Christophe Vergez}
\affiliation{Aix Marseille Univ, CNRS, Centrale Marseille, LMA, Marseille, France}
\author{Michaël Jousserand}
\affiliation{Buffet Crampon, 5 Rue Maurice Berteaux, 78711 Mantes-la-Ville, France}

\begin{abstract}
This study investigates the role of localized nonlinear losses in the register hole on the production of second-register notes.
First, an experiment is conducted to study the ability of a register hole to produce second register. 
A cylindrical tube is drilled with holes of increasing diameter.
Five are at the same level as the register hole of a B-flat clarinet, and five are at the same level as the thumb hole. 
Participant clarinetists  are then asked to play with constant control parameters.
At the beginning of each measurement, all holes are closed. 
The operator then opens randomly one of the ten holes.
The resulting register is noted.
The experiment is replicated numerically by time integration of two different models.
The first is the state-of-the-art model based on the modal decomposition of the input impedance of the resonator.
The second accounts for localized nonlinear losses in the register hole, through the model from Dalmont and Nederveen (2002).
These losses are handled through a variable modal coefficients method.
For the first model, simulations never produce second register, for any of the open holes.
For the second, the proportion of second-register production is close to the experiment for upstream holes, but remains at zero for downstream holes.
\end{abstract}
\maketitle

\section{Introduction}

A register hole is a side hole which, when opened, changes the oscillation frequency of a system from a fundamental frequency close to one mode of a resonator to a fundamental frequency close to a higher mode.
On the clarinet, where the frequencies of the higher modes are odd multiples of the frequency of the first mode, opening the register hole (by operating the chalumeau key, or twelfth key, or register key) enables the player to switch from the first register to the second register by a twelfth interval.

The clarinet register hole has mainly been studied by \citet{debut_analysis_2005} in a context of the optimization \citep{debut2003analysis} of the harmonicity between the first and second input impedance peaks.
The authors analytically determined the influence of the diameter, length and position of the hole on the frequencies of these two peaks. 
The role of the register hole in the production of second register notes was not explored. 
More recently, \citet{takahashi2014mode} have investigated the role of the register hole in the selection of the register produced at the emergence of the oscillation.
In this study, the resonator was modeled by a lossless  two-delayed system: one delay for the distance between the reed and the hole, the other delay to take into account the length of the main tube.
An operating region of the second-register production was determined, with respect to the ratio between the two delays (i.e. the relative position of the hole along the main tube), and the ratio between the amplitude of the first two peaks of the reflection function.
These results were not compared to a real instrument, for which the reflection function is more complex due to multiple open and closed holes bringing additional delays to the model.

For the clarinetist, the action of the register key from the low E$_2$\footnote{\label{note:1} All notes in this article are expressed in B$\,\flat$, as they would be written on a clarinet chart. A written E$_2$ corresponds to a heard D$_2$ in concert pitch (i.e. 146.8~Hz).}
to the F$_3$ of the left hand is straightforward.
Opening the register hole is a reliable method to play in the clarion (i.e. second) register.
However, in the framework of sound synthesis based on modal decomposition of input impedance measurements, the production of second-register notes is not guaranteed for second register fingerings.
In particular, it is possible for the physical model to produce a stable first register, whereas this situation is rarely, if ever observed on a real instrument.
This difference is problematic in the context of a sound synthesis intended to translate the behavior of a real instrument.
In an attempt to better reproduce the emergence of second register notes, the role of localized nonlinear losses in the register hole is investigated.

The input impedance is a linear quantity, expressing the frequency response of the resonator to a low-amplitude excitation. 
However, in the case of clarinet playing, the amplitude of the acoustic pressure can be very high:  \citet{backus1961vibrations} measured an acoustic pressure level of 160~dB inside the instrument for a soft tone, and 166~dB for a loud tone.
At a geometric discontinuity such as a lateral hole, different flow regimes are observed depending on the acoustic velocity amplitude \citep{ingaard1950acoustic}, ranging from laminar flow to vortex distachement \citep{martinez2023generation}.
Part of the jet's kinetic energy is absorbed by these vortices and dissipated as heat through friction, as observed by \citet[p. 217]{rayleigh1894theory} with a K\"onig resonator strongly excited by a tuning fork at its mouth: ``The formation of jet must take a serious draft on the energy motion".
These nonlinear losses were measured by \citet{ingaard1950acoustic} under the form of a purely real ``nonlinear impedance", such that the link between pressure and acoustic flow remains valid when nonlinear losses are taken into account. 
This nonlinear impedance depends on the acoustic velocity amplitude at the geometric discontinuity. 
In other ways, \citet{disselhorst1980flow} have developed a model of quasi-stationary flow for the boundary condition at the discontinuity at large amplitudes of acoustic velocity.

These two ways of modeling nonlinear losses have been extended to musical acoustics.
\citet{dalmont_experimental_2002} measured the evolution of the real part of the series and shunt impedances characterizing a side hole at high acoustic amplitudes. 
They developed a nonlinear impedance model, which is proportional to the amplitude of the acoustic velocity at the level of the hole, and to a coefficient linked to the roundness of its edges.
This work was continued by \citet{atig2004termination}, for a discontinuity at the end of a tube.

The quasi-stationary flow boundary condition has been applied to sound synthesis by waveguides \citep{ducasse1990modelisation, taillard2018phd} and delay lines \citep{atig2004saturation,guillemain2006digital,terroir2006}.
The inclusion of localized nonlinear losses in the physical models of woodwinds induces modifications to their dynamical behavior.
At the end of a tube, \citet{atig2004saturation} and \citet{dalmont_oscillation_2007} highlighted that nonlinear losses reduce the extinction threshold of the clarinet, i.e. the maximum blowing pressure that still produces an oscillating regime. 
In a side hole, \citet{keefe1983acoustic} noted a greater difficulty for a clarinetist to maintain a stable first register when the contribution of the nonlinear sound field was higher.
Nonlinear losses localized in side holes are also responsible for a decrease of the playing frequency, as shown by \citet[Appendix B]{debut_analysis_2005} 
 and by \citet[Chap. 3.2.2]{terroir2006} during a hole closing.
Finally, localized nonlinear losses in a side hole reduce the amplitude of the radiated acoustic pressure \citep[Chap. 3.2.3]{terroir2006} and could favor the amplitude of the second harmonic of the acoustic pressure spectrum \citep{keefe1983acoustic, guillemain2006digital}.
This last influence was not found by the authors in their last communication \citep{szwarcberg2023FA}.

In view of the wide variety of modifications introduced by localized nonlinear losses on the dynamics of a woodwind, their responsibility in the production of second-register notes is explored.
An experiment is first carried out with clarinetist participants blowing into a simplified clarinet made of a cylindrical tube. 
The tube is drilled with five holes of increasing diameter at the same position as a clarinet’s register hole, as well as five holes at the level of the left thumb hole. 
The clarinetist is blindfolded and asked to blow into the tube with constant blowing pressure and steady embouchure, while the operator opens one of the ten holes. 
The resulting register is noted.
The experiment is then replicated numerically, by time integration of a clarinet-type system of equations. 
Two different models are tested. 
The first is the standard model based on the modal decomposition of the input impedance of the resonator. 
The second accounts for localized nonlinear losses in the register hole, through the nonlinear losses model for side holes from \citet{dalmont_experimental_2002}. 
These losses are integrated into the physical model using a variable modal coefficients method \citep{diab2022nonlinear,szwarcberg2023amplitude}.
The experimental results are finally compared to the simulations, with and without nonlinear losses.
The discrepancies between the model and the experimental results are discussed.


\section{Experimental setup}
\subsection{Clarinet-like resonators}
The resonator considered in this study was first manufactured by \citet{petersen_link_2020} and is made of a cylindrical tube of length 450~mm, diameter 13~mm and first resonance frequency 185~Hz.  
A clarinet player hears the note F$_3$\textsuperscript{\ref{note:1}} when playing the tube.
In comparison, a B\,$\flat$ clarinet has a length of 588~mm and its lowest note is E$_3$.
Ten holes have been drilled in two positions along the axis of the resonator, labeled ``upstream'' (U) and ``downstream'' (D).
The upstream position corresponds to the same position as the register hole of a real clarinet: 84~mm from the top of the barrel.
The downstream position is at the same location as the hole of the left hand thumb (F$_4$): 183~mm from the top of the barrel.
Five holes are drilled at each position, with diameters of 1.0~mm, 1.5~mm, 2.4~mm, 3.0~mm, and 5.0~mm.
In comparison, the diameters of the register hole and of the left thumb hole of the clarinet measure approximately 2.5~mm and 7.5~mm respectively.
They are represented on Figure~\ref{pic:1}, covered of striped tape. 
The input impedance of the prototype is then measured, for all holes closed, and each of the 10 holes open successively.
Each measurement is repeated twice. 
A cylindrical adaptation piece, 16~mm in diameter and 65~mm long (similar volume to a clarinet mouthpiece), connects the impedance sensor to the prototype.
Measured impedances are plotted in dotted lines on Figure~\ref{fig:2}.

\subsection{Playing tests}

\begin{figure}[h!]
	\centering
	\includegraphics[width=.45\textwidth]{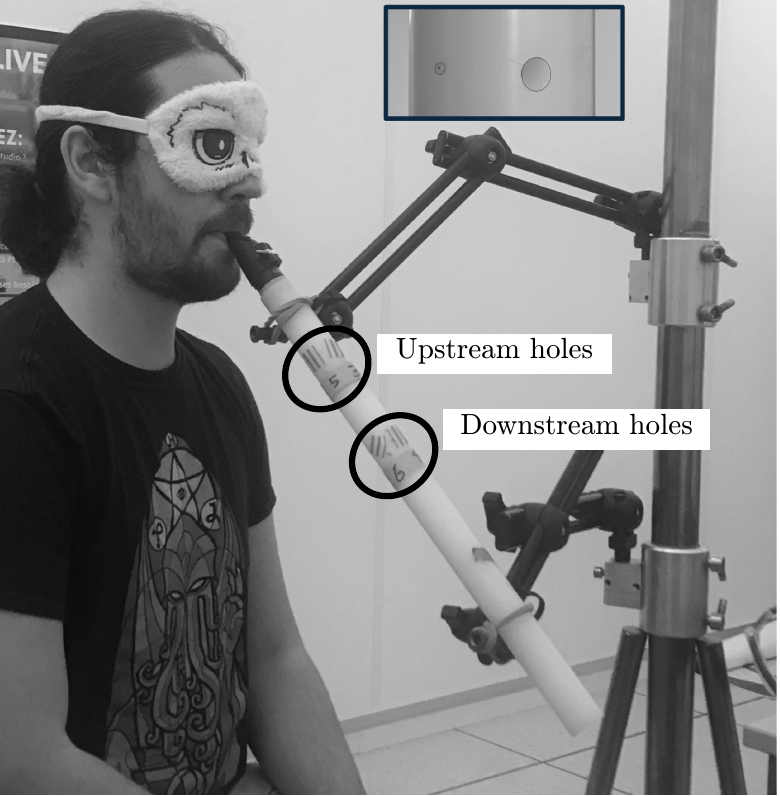}
	\caption{Photography of the experimental setup.}
	\label{pic:1}
\end{figure}

Playing tests were carried out to assess the ability of each of the ten side holes to produce second-register whey they open. 
Fourteen clarinetists of different levels (from very beginner to graduate clarinetist) participated to the experiment.
Each musician was allowed to use their own equipment (Vandoren 5RV, BD5 and M30 mouthpieces, Vandoren reeds of strength 3.0 to 3.5).
Before the test, the participant installs his or her equipment on the prototype, all holes being covered with adhesive tape. 
The participant adjusts the instrument to his or her preference on a stand, in order to blow comfortably into the prototype (see Figure \ref{pic:1}). 
During the test, the only contact zone between the musician and the instrument is at the mouthpiece: the hands do not touch the instrument. 
The clarinetist is blindfolded throughout the test.

The playing test proceeds as follows. First, the clarinetist is seated blindfolded, ready to play. 
The operator removes the tape from one of the ten side holes. 
He then closes the hole with his finger.
The clarinetist then blows into the instrument to play the note corresponding to the first register of the closed prototype (note F$_2$).
The clarinetist is asked to perform a ``long tone", i.e. to play while maintaining a constant pressure and embouchure, in order to achieve with ease a sound perceived as pleasant.
While the musician blows, the operator opens the hole.
The resulting register or behavior is noted, in addition to being recorded by a microphone.
The operator then covers the hole again with adhesive tape, and moves on to the next hole.

During a test session, each hole is evaluated twice. 
The hole to be tested is chosen randomly.
There are therefore 20 random draws without replacement. 
Each participant completes three sessions. 
The first is a training session, the results of which are not recorded.
The complete test lasts approximately thirty minutes.

\section{Model}
The experiment previously described is reproduced by numerical simulations. 
First, a digital model of the simplified clarinet is developed.
The geometry of the numerical resonator is optimized according to the measured input impedance.
The method enabling to account for localized nonlinear losses in the register hole within the modal decomposition of the input impedance is then detailed.
The equations for the self-sustained oscillations of the clarinet-like system are finally presented.

\subsection{Numerical model of the resonator}
\subsubsection{Transfer matrices}
\begin{figure}[h!]
	\centering
	\includegraphics[width=.49\textwidth]{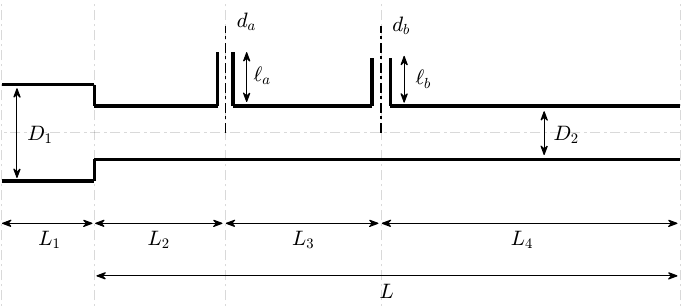}
	\caption{Scheme of the digital model of simplified resonator.}
	\label{fig:1}
\end{figure}

Each resonator is modeled by the Transfer Matrix Method from a simplified model, whose dimensions are specified in Figure~\ref{fig:1}.
The model consists of four cylinders with respective dimensions $\{L_1,D_1\}$, $\{L_2,D_2\}$, $\{L_3,D_2\}$, and $\{L_4,D_2\}$.
The transfer matrix of a cylinder of length $L$ is given by:
\begin{equation}
M_{c} =\left( \begin{matrix}
\cos(\bar{k} L) & j Z_c \sin(\bar{k} L) \\
j Z_c^{-1} \sin(\bar{k} L) & \cos(\bar{k} L)
\end{matrix}
\right),
\end{equation}
where $Z_c=\sqrt{Z_v/Y_t}$ is the characteristic impedance of plane waves within the tube, and $\bar{k}=-j\sqrt{Z_v Y_t}$ is the complex wave number, accounting for viscous ($Z_v$) and thermal ($Y_t$) dissipation.
Definitions of $Z_v$ and $Y_t$ are accessible through \citet[Chap. 5.5]{bible2016}.

A cross-section change transfer matrix $M_s$ links the tube $\{L_1,D_1\}$ to the tube $\{L_2,D_2\}$:
\begin{equation}
M_{s} =\left( \begin{matrix}
1 & j \omega m_d \\ 0 & 1 
\end{matrix} \right),
\end{equation}
where $\omega$ is the angular frequency and $m_d$ is the added mass linked to the cross-section change \citep[Eqs.  (7.158) and (7.162)]{bible2016}.

Two lateral holes of dimensions $\{$height, diameter$\}$  labeled respectively $\{\ell_a,d_a\}$ and $\{\ell_b,d_b\}$ separate the tubes of lengths $L_2$, $L_3$ and $L_4$. 
A side hole is represented by a T-circuit with two equal series impedances $Z_a/2$ and a parallel impedance $Z_s$, as described by \citet{garcia_mayen_characterization_2021, dalmont_experimental_2002}.
The transfer matrix of the hole $M_h$ is defined as:
\begin{equation}\label{eq:Mh}
M_h = \frac{1}{1- \dfrac{Z_a}{4 Z_s}} \left( \begin{matrix}
1 + \dfrac{Z_a}{4 Z_s} & Z_a \\ \dfrac{1}{Z_s} & 1 + \dfrac{Z_a}{4 Z_s}
\end{matrix} \right).
\end{equation}
The series impedances $Z_a$ are defined by:
\begin{equation}
Z_a = j Z_c k t_a,
\end{equation}
where $k=\omega/c_0$ is the wavenumber, and $t_a$ the series length correction \cite[Eq. (7)]{garcia_mayen_characterization_2021}, depending on the chimney length and on the cross-section ratio between the chimney and the main bore.
For a closed hole, the parallel impedance $Z_s^{(c)}$ is \citep[Eq. (7)]{nederveen1998corrections}:
\begin{equation}
Z_s^{(c)} = j Z_{ch} \left(\bar{k}_h t_i -1 /  \tan \left[\bar{k}_h (\ell + t_m ) \right] \right),
\end{equation}
where $Z_{ch}$ is the characteristic impedance of the hole, $\bar{k}_h$ is the complex wavenumber within the hole, $\ell$ is the length of the hole chimney, $t_i$ is a length correction due to evanescent modes \citep[Eq. (8)]{garcia_mayen_characterization_2021} and $t_m$ is the length correction due to the matching volume \citep[Eq. (37)]{nederveen1998corrections}.
For an open hole, the parallel impedance $Z_s^{(o)}$ is written as \citep[Eq. (5)]{garcia_mayen_characterization_2021}:
\begin{equation}
Z_s^{(o)} = j Z_{ch} \left( k t_i + \tan \left[ \bar{k}_h \ell + k (t_m + t_R ) \right] \right),
\end{equation}
where $t_R=Z_{Rh}/(j k Z_{ch})$ is the (complex) length correction due to the radiation of the open hole, of radiation impedance $Z_{Rh}$ \cite[Eq. (20)]{silva_approximation_2009}.
Lateral holes are supposed to be infinitely flanged.
The open end of the tube, of radiation impedance $Z_R$, is supposed to be unflanged.

Following the scheme in Figure~\ref{fig:1}, the complete transfer matrix of the resonator is denoted $M_{\mathrm{tot}}$, such that:
\begin{equation} \label{eq:Mtot}
M_\mathrm{tot} = M_{c1} M_s M_{c2} M_{ha} M_{c3} M_{hb} M_{c4}.
\end{equation}
Matrices $M_{ha}$ and $M_{hb}$ can be either in ``open hole" (superscript $^{(o)}$) or ``closed hole" (superscript $^{(c)}$) configuration.
From the coefficients of $M_\mathrm{tot}$, the input impedance $Z_\mathrm{in}$ is defined: 
\begin{equation}\label{eq:Zin}
Z_\mathrm{in} = \frac{M_\mathrm{tot}^{(1,1)} Z_R + M_\mathrm{tot}^{(1,2)}}{M_\mathrm{tot}^{(2,1)} Z_R + M_\mathrm{tot}^{(2,2)}},
\end{equation}
and its dimensionless equivalent: $z_\mathrm{in} = Z_\mathrm{in}/Z_c$.

\subsubsection{Fitted impedances}

Measuring on the prototype the 10 geometric parameters shown in Figure~\ref{fig:1} would constitue the ``direct modeling approach" to compute $Z_\mathrm{in}$ using Eq.~\eqref{eq:Zin}.
However, since we are interested in proposing the numerical model that best fits the measured input impedances, an ``inverse modeling approach" is followed where the ten geometric parameters are determined through optimization.

\begin{figure}[h!]
	\centering	
	\includegraphics[width=.45\textwidth]{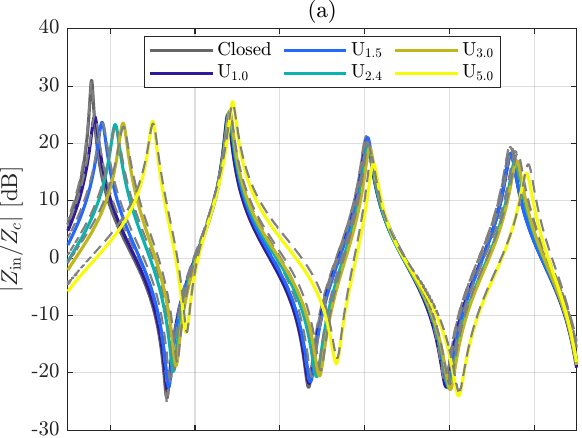}
	\includegraphics[width=.45\textwidth]{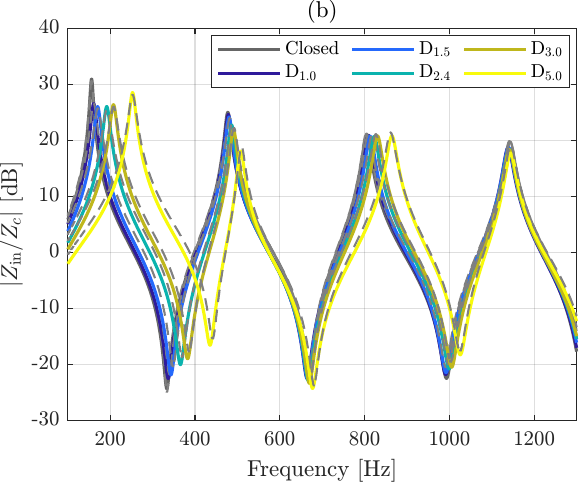}	
	\caption{Fitted impedances of the resonator, when the hole is open upstream (a) and downstream (b). Curves in dashed lines correspond to the measurements. The open hole is designated by U$_n$ or D$_n$, (with $n$ the value of the hole diameter in mm) for upstream and downstream positions respectively.}
	\label{fig:2}
\end{figure}

In the context of modal decomposition, only impedance peaks are considered: it seems therefore particularly relevant to fit their amplitude and frequency.
Several cost functions focusing on the fitting of the impedance peaks were tested, following \citet{colinot2019numerical} and  \citet{ernoult2020woodwind}.
The following  cost function was finally selected:
\begin{equation}
J(\chi ) =\sum_{\omega_i=\omega_\mathrm{min}}^{\omega_\mathrm{max}} \| \, |z_\mathrm{in}^\odot(\omega_i)| - |z_\mathrm{in}(\omega_i,\chi)| \, \|^p,
\end{equation}
where $\chi$ is the vector of the geometric quantities to be optimized, $p\in\mathbb{N}$, and $z_\mathrm{in}^\odot$ is the dimensionless impedance measured between 100~Hz and 4000~Hz with a step size of 0.2~Hz using the CTTM impedance sensor \citep{dalmont2008new}.
Increasing the exponent $p$ results in better matching of high-amplitude peaks at the expense of low-amplitude peaks, as depicted by \citet{colinot2019numerical}.
After several trials, the value $p=8$ was selected for each impedance, except for the open holes U$_{5.0}$ and D$_{5.0}$ where $p=4$ was chosen.
Optimization results are noted in Table~\ref{tab:1} and shown in Figure~\ref{fig:2}.

\begin{table}[ht!]
\caption{Dimensions of digital resonator geometrical parameters after optimization, in millimeters. 
The labels refer to the position of the open hole (\textbf{U}p, \textbf{D}own or \textbf{C}losed), as well as their hole diameter in mm.
Hole diameters written in bold indicate that the corresponding hole is open.  }
\label{tab:1}
\begin{ruledtabular}
\begin{tabular}{lcccccccccc} 
Label & $L_1$ & $D_1$ &$L_2$ & $D_2$ & $L_3$ & $L_4$ & $\ell_a$ & $d_a$ & $\ell_b$ & $d_b$ \\
\hline
C & 64.7 & 12.9 & 80.0 & 11.3 & 80.3 & 292 & 1.0 & 3.2 & 5.5 & 6.2 \\
U$_{1.0}$ & 66.7 & 12.8 & 55.9 & 11.2 & 100 & 296 & 10.0 & \textbf{0.76} & 11.2 & 5.0\\
U$_{1.5}$ & 64.8 & 12.8 & 47.5 & 11.4 & 72.1 & 334 & 16.2 & \textbf{1.5} & 6.3 & 2.0 \\ 
U$_{2.4}$ & 66.7 & 12.4 & 62.6 & 11.0 & 94.8 & 294 & 16.0 & \textbf{2.4} & 4.3 & 6.6 \\ 
U$_{3.0}$ & 67.4 & 12.2 & 63.3 & 11.0 & 83.6 & 304 & 16.9 & \textbf{3.0} & 4.9 & 3.4 \\ 
U$_{5.0}$ & 68.4 & 11.8 & 63.1 & 10.4 & 36.7 & 350 & 16.3 & \textbf{4.6} & 1.0 & 2.0 \\ 
D$_{1.0}$ & 65.0 & 12.8 & 70.1 & 11.4 & 74.9 & 309 & 7.5 & 5.8 & 13.1 & \textbf{0.82} \\
D$_{1.5}$ & 64.6 & 12.6 & 69.4 & 11.2 & 75.6 & 309 & 3.8 & 7.8 & 22.5 & \textbf{1.6} \\
D$_{2.4}$ & 65.6 & 12.4 & 68.2 & 11.0 & 93.3 & 290 & 12.4 & 4.8 & 19.4 & \textbf{2.6} \\
D$_{3.0}$ & 65.7 & 12.6 & 70.0 & 11.4 & 92.8 & 289 & 9.1 & 5.6 & 13.8 & \textbf{2.8} \\
D$_{5.0}$ & 66.3 & 12.0 & 65.4 & 10.8 & 97.2 & 288 & 13.1 & 4.6 & 11.0 & \textbf{5.0} \\
\end{tabular}
\end{ruledtabular}
\end{table}

\subsection{Accounting for localized nonlinear losses through modal coefficients}

Nonlinear losses were measured by \citet{ingaard1950acoustic} under the form of a purely real ``nonlinear impedance" (such that the link between pressure and acoustic flow remains valid when nonlinear losses are taken into account), depending on the acoustic velocity amplitude at the geometric discontinuity. 
It also depends on the roundness of the edges \citep{thurston1958nonlinear}, through a loss coefficient whose values have been measured and tabulated by \citet{dalmont_experimental_2002} for a lateral hole and by \citet{atig2004termination} for an open cylindrical tube.
In the model presented below, the ``nonlinear impedance" accounting for losses in the register hole is incorporated into the input impedance of the resonator. 
A variable modal coefficients method based on the work of \citet{diab2022nonlinear} enables to take into account localized nonlinear losses in the system of equations for the self-sustained oscillations of a clarinet.

\subsubsection{Nonlinear losses model for the open side hole}
\citet{dalmont_experimental_2002} measured the impedance evolution of open lateral holes of different diameter and roundness, with respect to the acoustic velocity amplitude in the hole $v_\mathrm{RMS}$.
According to their experimental results, the real part of the series impedance $Z_a^{(o)}$ increases with $v_\mathrm{RMS}$ from the following linear relationship: 
\begin{equation}
Z_a^\mathrm{(NL)} = Z_a^{(o)} + K_a Z_c v_\mathrm{RMS}/c_0,
\end{equation}
where $K_a=0.4$ for a hole with sharp edges.
The real part of the parallel impedance $Z_s^{(o)}$ increases according to the equation :
\begin{equation}
Z_s^\mathrm{(NL)} = Z_s^{(o)} + \left( K_h Z_{ch} + K_a Z_c /4 \right) v_\mathrm{RMS}/c_0,
\end{equation}
with $K_h=0.5$ for a sharp-edged hole drilled in a PVC tube \citep{dalmont_experimental_2002}.  

\subsubsection{Computation of the modal coefficients}\label{sec:modalCoeffs}
The modal decomposition of the input impedance into poles and residues enables to approximate $Z_\mathrm{in}$ by a sum of $N$ complex modes, according to the following expression:
\begin{equation}\label{eq:Zn}
Z_\mathrm{in} = Z_c \sum_{n=1}^N \frac{C_n}{j \omega - s_n} + \frac{C_n^*}{j \omega - s_n^*},
\end{equation}
where the superscript $^*$ refers to the complex conjugate.
In this study, the number of modes is limited to $N=12$ since the impedance measured only exhibit the twelve first peaks.
To account for localized nonlinear losses in the open hole through modal decomposition simulation, the residues $C_n$ and poles $s_n$ are now considered to depend on $v_\mathrm{RMS}$ \citep{diab2022nonlinear}.
To do this, we first replace the impedances $Z_a$ and $Z_s^{(o)}$ by $Z_a^\mathrm{(NL)}$ and $Z_s^\mathrm{(NL)}$ in the open hole transfer matrix $M_h$ (Eq. \eqref{eq:Mh}). 
The transfer matrix of the resonator $M_\mathrm{tot}$ is then obtained, followed by the expression of the input impedance (Eq. \eqref{eq:Zin}) with respect to both the frequency and $v_\mathrm{RMS}$.
Noting $s$ the Laplace variable, the poles $s_n(v_\mathrm{RMS})$ are defined by the values of $s$ for which the denominator of $Z_\mathrm{in}$ becomes zero. 
Furthermore, writing $Z_\mathrm{in} = \mathcal{N}(s, v_\mathrm{RMS})/ \mathcal{D}(s, v_\mathrm{RMS})$ (and ensuring that the numerator $\mathcal{N}$ has no pole), the poles $s_n$ are given by:
\begin{equation}\label{eq:sn}
\mathcal{D}(s_n, v_\mathrm{RMS})=0.
\end{equation}
This equation is solved numerically.
Application of the residues theorem then gives the residue expression $C_n(v_\mathrm{RMS})$ \citep[Sec. 3.1.2]{szwarcberg2023amplitude} :
\begin{equation}\label{eq:Cn}
C_n(v_\mathrm{RMS}) = \dfrac{\mathcal{N}(s_n, v_\mathrm{RMS}) }{Z_c \dfrac{\partial \mathcal{D}}{\partial s} (s_n, v_\mathrm{RMS})}.
\end{equation}

\begin{figure*}
\centering
\includegraphics[width=\textwidth]{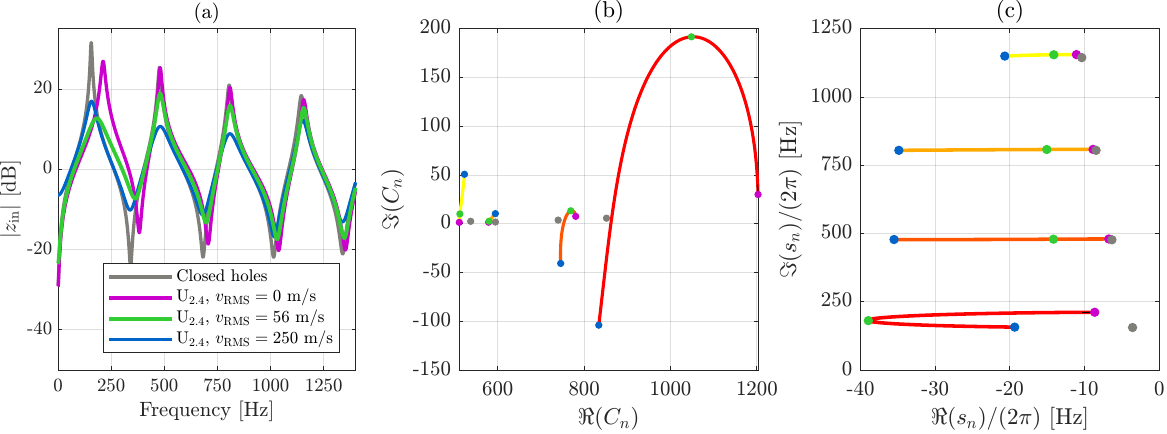}
\caption{Evolution of the input impedance (a) of resonator U$_{2.4}$, and of its 4 first complex residues (b) and poles (c), with respect to $v_\mathrm{RMS}$. 
The colors of the modal coefficients range from red (mode 1) to yellow (mode 4). 
The evolution of the modal coefficients is represented for $v_\mathrm{RMS}\in [0,250]$~m/s. The modal coefficients computed at the same acoustic velocities as the input impedances of resonator U$_{2.4}$ are indicated by dots in the same color as the curves.}
\label{fig:3}
\end{figure*}

Figure \ref{fig:3}(a) shows the evolution of the input impedance of resonator U$_{2.4}$ for three different values of $v_\mathrm{RMS}$.
We first focus on all peaks except the first.
The amplitude of these peaks decreases as nonlinear losses increase, as reported by \citet{keefe1983acoustic}: ``Nonlinear losses lower the peak heights".
Anti-resonances are also less pronounced as nonlinear losses increase.
Figure \ref{fig:3}(c) presents a monotonic decrease of the $\Re(s_n)$ for $n\geq2$ when $v_\mathrm{RMS}$ increases. 
However, there is no common trend for the complex residues $C_n$ for $n\geq2$, as shown on Figure \ref{fig:3}(b).
Another point of view can be adopted by switching to real modes of amplitudes $A_n$ and $B_n$, damping ratios $\xi_n$ and natural frequencies $\omega_n$.
Equivalences are provided by \citet{ablitzer2021peak}, such that:
\begin{align*}
A_n &= 2 \Re(C_n), & B_n &= 2 \Re(C_n s_n^*),\\
\xi_n &= \frac{1}{\sqrt{1 + \left[\Im(s_n)/\Re(s_n)\right]^2}}, & \omega_n &= - \Re(s_n)/\xi_n.
\end{align*}
For modes of index $n\geq2$, the amplitudes $A_n$ do not follow a common trend when $v_\mathrm{RMS}$ increases.
The same is observed for $B_n$ and $\omega_n$.
However, $\xi_n$ increases almost linearly when nonlinear losses increase. 
Moreover, the higher the index of the mode, the lower the slope of $\xi_n(v_\mathrm{RMS})$.
The increase of the modal damping ratios is also observed in simulations accounting for localized nonlinear losses at the open end of a tube \citep[Fig. 1]{szwarcberg2023FA}.

The first impedance peak evolves differently from the others when nonlinear losses increase.  
In comparison with the other peaks, its frequency  decreases strongly: Figure \ref{fig:3}(c) shows a monotonic decrease in $\Im(s_1)$ as $v_\mathrm{RMS}$ increases.
Furthermore, the amplitude of the first impedance peak decreases for $v_\mathrm{RMS}\in [0,56]$~m$/$s to a minimum.
It is indicated by the extrema of $\Im(C_1)$ and $\Re(s_1)$, represented by a green dot on Figures \ref{fig:3}(b) and \ref{fig:3}(c).
From the perspective of real modes, this extremum is only found for $\xi_1$, which is maximal around $v_\mathrm{RMS}=56$~m$/$s.
For $v_\mathrm{RMS}\geq 56$~m$/$s, the first peak regains amplitude, which is reflected in a diminution of the damping ratio, balanced by a diminution of $A_1$ and $B_1$.
The shape of the first impedance peak becomes progressively similar to that of the closed hole.
This behavior is also reported by \citet[Appendix B]{debut_analysis_2005}: when the resistive part of the open tone hole impedance tends to infinity, the frequency of the first input impedance peak tends to the frequency of the first peak of the same resonator with a closed hole.
However, the amplitude of the first impedance peak does not become as high as that of the closed hole: even for unrealistically high values of $v_\mathrm{RMS}$ (250~m$/$s and above), its amplitude remains smaller than 20~\% of the closed hole's first impedance peak.

Finally, the crossing of the real axis by $C_1$ and $C_2$ as nonlinear losses increase raises questions about the passivity of the input impedance. 
First, it was numerically checked that the input impedances expressed under their analytical form (Eq.~\eqref{eq:Zin}) all verify $\Re \left[ Z_\mathrm{in}(j \omega, v_\mathrm{RMS})\right] \geq 0$ for all $\omega>0$ and $v_\mathrm{RMS}>0$.
The modal decomposition (Eq.~\eqref{eq:Zn}) respects this condition, provided that a sufficient number of modes is accounted for.
For 12 modes, passivity is ensured for all resonators except U$_\mathrm{2.4}$, U$_\mathrm{3.0}$ and U$_\mathrm{5.0}$.

\subsubsection{Fitting of the modal coefficients}
\begin{figure}[h!]
	\centering
	\includegraphics[width=.49\textwidth]{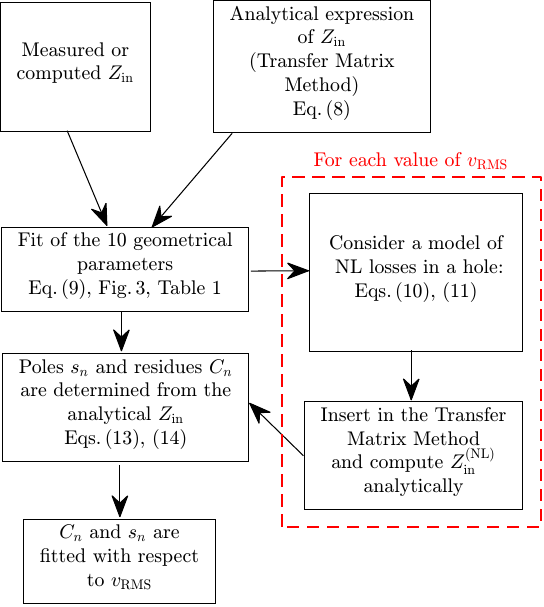}
	\caption{Summary scheme of the model of modal decomposition of the input impedance accounting for localized nonlinear losses in a side hole.}
	\label{fig:method}
\end{figure}

In a time integration simulation, the modal coefficients are recomputed at each iteration of the time integration scheme, depending on the value of the acoustic velocity amplitude computed in the open hole.
However, solving Eq.~\eqref{eq:sn} to calculate $s_n(v_\mathrm{RMS})$ takes a few seconds on a standard computer.
This is not compatible with a simulation intended to be as close as possible to real time.
One solution is to fit the real and imaginary parts of the modal coefficients with respect to $v_\mathrm{RMS}$.
Following the method of \citet{diab2022nonlinear}, the modal coefficients were fitted as rational fractions by Vector Fitting \citep{gustavsen1999rational} using the Matlab routine \texttt{vecfit3.m} \cite{gustavsen2006improving, deschrijver2008macromodeling}.
The real and imaginary parts of the modal coefficients are fitted separately.
The fitted coefficients are thus denoted $\tilde{C}_n(v_\mathrm{RMS})$ and $\tilde{s}_n(v_\mathrm{RMS})$, such that
\begin{equation}
\tilde{C}_n(v_\mathrm{RMS}) = \tilde{C}_n^\Re(v_\mathrm{RMS}) + j \tilde{C}_n^\Im(v_\mathrm{RMS}).
\end{equation}
The same principle applies for $\tilde{s}_n$.
The presented method is summarized on Figure~\ref{fig:method}.

\subsubsection{Mode shapes of the resonator}
Calculating the acoustic velocity in the open hole requires first to know the acoustic pressure $p_h$ in the hole, and therefore the mode shapes at the location of the hole $\phi_n(x_h)$ linking $p_h$ to the modal pressures at the input $p_n$ (Eq.~\eqref{eq:phphin}).
These mode shapes are computed using the Transfer Matrix formulation of the resonator. 
First, the product of the transfer matrices corresponding to the parts of the resonator located upstream from the open hole is denoted $M_u$. 
By isolating the first series branch of the T-circuit of $M_h^{(o)}$ (see, for instance, Fig. 7.25 from \citet{bible2016}), the link between the acoustic field at the tube input and in the hole is given by:
\begin{equation}
\begin{pmatrix}
P_\mathrm{in} \\ U_\mathrm{in}
\end{pmatrix} = 
M_u \begin{pmatrix}
1 & Z_a^{(o)}/2 \\ 0 & 1
\end{pmatrix}
\begin{pmatrix}
P_h \\ U_L
\end{pmatrix},
\end{equation}
in the frequency domain.
$U_L$ refers to the acoustic flow in the left branch of the T-circuit.
Isolating $P_h$, then bringing out the explicit expression of $Z_\mathrm{in}$ (Eq. \eqref{eq:Zin}) yields:
\begin{equation}
P_h = U_\mathrm{in} \left[ D_h Z_\mathrm{in} - B_h \right],
\end{equation}
where 
\begin{align*}
B_h&= M_u^{(1,1)} Z_a^{(o)}/2 + M_u^{(1,2)}  \quad \mathrm{and} \\
D_h &= M_u^{(2,1)}Z_a^{(o)}/2 + M_u^{(2,2)}.
\end{align*}
Since $Z_\mathrm{in}= \mathcal{N}/ \mathcal{D}$, gathering the terms to the same denominator gives then:
\begin{equation}\label{eq:phiDenom}
P_h = U_\mathrm{in}\frac{D_h \mathcal{N} - B_h \mathcal{D}}{ \mathcal{D}}.
\end{equation}
The poles of Eq. \eqref{eq:phiDenom} are the same poles as the poles of $Z_\mathrm{in}$, computed through Eq.~\eqref{eq:sn}.
Following the method of \citet[Chap. 5.6.3, Eq. (5.175)]{bible2016}, the zero-order expansion of the numerator terms around $s_n$ enables to write the expansion of $P_h$:
\begin{equation}
P_{h, n} = U_\mathrm{in}\frac{D_h(s_n) \mathcal{N}(s_n) }{(s-s_n) \dfrac{\partial \mathcal{D}}{\partial s} (s_n)}.
\end{equation}
The identification of the series expansion of $P_\mathrm{in}$ in the previous equation gives the following relation:
\begin{equation}
P_{h, n} = D_h(s_n) P_{\mathrm{in}, n}.
\end{equation}
In the time domain this finally leads to:
\begin{equation}\label{eq:phi}
\phi_n(x_h) =\Re \left[ M_u^{(2,1)}(s_n)\frac{Z_a^{(o)}(s_n, v_\mathrm{RMS})}{2} + M_u^{(2,2)}(s_n) \right].
\end{equation}

Note that in Eq.~\eqref{eq:phi}, the imaginary part is ignored.
The imaginary part of the mode shapes reflects the coupling between modes due to damping, as explained by \citet[Chap. 5.2.1.1]{bible2016}.
Since the eigenfrequencies of the conservative equivalent of the resonator are ``sufficiently well separated" from each other, and supposing that the inter-modal damping coefficients are sufficiently weak, 
it is legitimate to consider that the eigenmodes remain unchanged compared to the conservative case, i.e. $\phi_n$ are purely real.

\begin{figure}[h!]
	\centering
	\includegraphics[width=.45\textwidth]{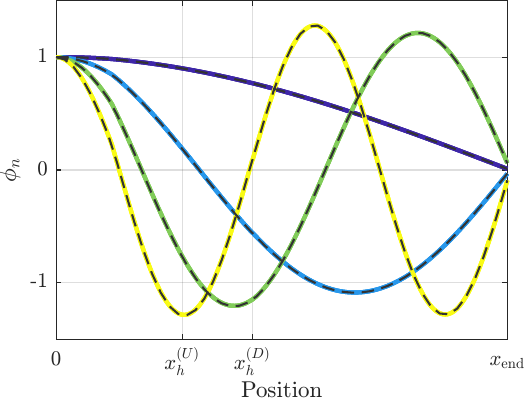}
	\caption{Evolution of the four first mode shapes of the resonator with all holes closed. Geometrical dimensions are given in the first row of Table~\ref{tab:1}.  Color of modes in ascending order: dark blue, light blue, green, yellow. Dashed lines: simulation with the Openwind software \citep{chabassier2020virtual}.}
	\label{fig:phi}
\end{figure}

The four first mode shapes of the resonator with all holes closed (Label C in Table~\ref{tab:1}) are represented on Figure~\ref{fig:phi}.
In the vicinity of $x_h^{(U)}$ (upstream hole), $\phi_2(x_h^{(U)})$ is close to zero, suggesting that the second modal pressure would be only slightly affected by the opening of the hole.
These observations are reflected on the impedance peaks of Figure~\ref{fig:2}(a): when the upstream hole is open, the amplitude of the second peak is only slightly attenuated.
If the four first modes except the second are attenuated, we expect to observe second-register regimes when the upstream hole is open.
The same applies for $\phi_4(x_h^{(D)})$ (downstream hole), the fourth impedance peak of Figure~\ref{fig:2}(b) and the emergence of the fourth register.

\subsection{Equations for the self-sustained oscillations}
Three main parts characterize the self-sustained oscillations of the clarinet-like system: reed dynamics, reed channel, and resonator.
The equations of the physical model are very similar to  \citet{szwarcberg2023amplitude}, with some adjustments for the computation of $v_\mathrm{RMS}$.

\subsubsection{Reed dynamics}
The reed is a single degree-of-freedom oscillator, of angular resonance frequency $\omega_r$ and damping $q_r$.
The dimensionless displacement $x(t)$ is given by:
\begin{equation}\label{eq:reed}
\dfrac{1}{\omega_r^2}\ddot{x}(t) +\dfrac{q_r}{\omega_r}\dot{x}(t)+ x(t)=p(t)- \gamma,
\end{equation}
where $p(t)$ is the dimensionless acoustic pressure at the input of the resonator, and $\gamma=P_m/p_M$ is the dimensionless blowing pressure.
The beating pressure $p_M=4$~kPa \citep{dalmont_oscillation_2007} is the pressure for which the blowing pressure $P_m$ closes the reed channel, in quasi-static regime. 
A ``ghost reed" model is considered \citep{colinot2019ghost}, meaning there is no contact force between the reed and the tip of the mouthpiece.
The control parameters $\gamma$, $\omega_r$ and $q_r$ remain constant in every simulations. 
Their values are detailed in Table~\ref{tab:2}.

\subsubsection{Reed channel}
The flow through the reed channel $u(t)$ is a dimensionless equivalent of the Bernoulli equation for steady flow assuming total turbulent dissipation \citep[Eq. (4)]{wilson_operating_1974}, including the flow induced by the motion of the reed \citep{dalmont2003nonlinear}:
\begin{equation}
u(t)=-\lambda \dot{x}(t) +  \zeta \left[x(t)+1\right]^+  
\mathrm{sgn} \bigl[\gamma-p(t)\bigr] \sqrt{\left|\gamma-p(t)\right|},
\end{equation}
where the superscript $^+$ denotes the positive part, $\lambda=5.5 \cdot 10^{-3}/c_0$ is the reed flow parameter \citep{chabassier2022control}, and $\zeta$ is the embouchure control parameter.

\subsubsection{Resonator}
When taking into account nonlinear losses in the register hole, the time-domain equivalent of Eq. \eqref{eq:Zn} is written as:
\begin{equation}\label{eq:pn}
\dot{p}_n(t) = \tilde{C}_n(v_\mathrm{RMS}) u(t) + \tilde{s}_n(v_\mathrm{RMS}) p_n(t),
\end{equation} 
where $p_n(t)$ are the modal acoustic pressures at the input of resonator. 
They are linked to the acoustic pressure $p(t)$ through the relation:
\begin{equation}
p(t)=2 \sum_{n=1}^N \Re\left[p_n(t)\right].
\end{equation}
In explicit time-domain simulation, the value of $v_\mathrm{RMS}$ is recomputed at each time step. 
The following section details how to compute the RMS acoustic velocity in the open side hole.

\subsubsection{RMS acoustic velocity in the open hole}
The first step in determining $v_\mathrm{RMS}$ is to calculate the sound pressure $p_h(t)$ in the hole.
It is calculated from the modal pressures $p_n(t)$ (Eq. \eqref{eq:pn}) and the mode shapes $\phi_n(x_h)$ (Eq. \eqref{eq:phi}) of the waveguide at the location of the hole $x_h$:
\begin{equation}\label{eq:phphin}
p_h(t) = 2 \sum_{n=1}^{N} \Re \left[ p_n(t) \right] \phi_n(x_h) .
\end{equation}
The acoustic velocity in the hole $v_h(t)$ is then computed.
Given that the fitting of the modal coefficients was realized with chosen dimensioned values of $v_\mathrm{RMS}$, the acoustic velocity $v_h(t)$ must be dimensioned.
It is computing following the model of a quasi-stationary oscillating flow from \citet{disselhorst1980flow} and the simplification to the unsteady case from \citet[Eq. (20)]{atig2004saturation}:
\begin{equation}
{p}_h(t) p_M= \frac{ \rho c_d}{2}  v_h^2(t) \, \mathrm{sgn}\left[p_h(t)\right],
\end{equation}
where $c_d=\dfrac{3 \pi}{4} K_h=1.18$ for $K_h=0.5$, and $p_h(t)p_M$ is the dimensioned equivalent of $p_h(t)$.
Hence,
\begin{equation}
v_h(t)=  \sqrt{\frac{2 p_M}{\rho c_d} |p_h(t)| } \, \mathrm{sgn}\left[p_h(t)\right].
\end{equation}
The RMS velocity is finally computed:
\begin{equation}
\frac{\partial(\tau v^2_\mathrm{RMS})}{\partial t} = v_h^2(t) - v^2_\mathrm{RMS}(t),
\end{equation}
where $\tau=\Im(s_1/(2\pi))^{-1}$ is a short-memory term used to avoid numerical divergence when integrating $v_h(t)$ \citep{szwarcberg2023amplitude}.

\section{Results}
\subsection{Playing tests}

\begin{figure}[h!]
	\centering
	\includegraphics[width=.45\textwidth]{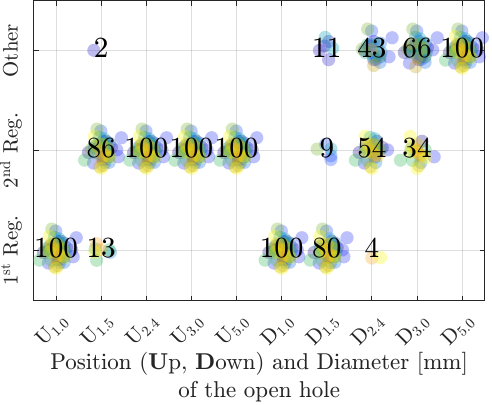}
	\caption{Experimental results: percentage of production of a specific register (first register, second register, or other) for each open hole. Each dot refers to one measurement. Each color refers to one participant.}
	\label{fig:resultsExp}
\end{figure}

\begin{figure*}[t]
	\centering
	\includegraphics[width=\textwidth]{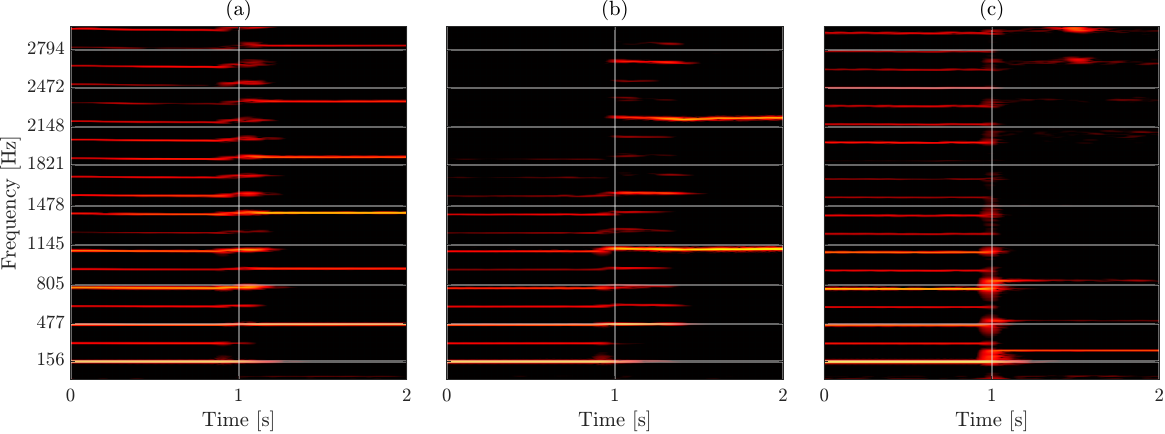}
	\caption{Spectrograms of some registers produced during the playing tests. Holes are open at $t_0=1$~s. The frequency axis ticks are resonance frequencies of the closed holes resonator.
	Panel (a): $2^\mathrm{nd}$ reg. after opening hole U$_{1.5}$. Panel (b): $4^\mathrm{th}$ reg. after opening hole D$_{5.0}$. Panel (c): ``muffled sound" after opening hole D$_{5.0}$.}
	\label{fig:spec}
\end{figure*}

The results from the playing tests are presented in Figure~\ref{fig:resultsExp}.
Each of the ten holes has been played 56 times (14 players $\times$ 2 sessions $\times$ 2 trials).
The sound produced has been classified from the instrument maker's point of view, i.e. whether  the hole is reliable to play twelfths.
Three categories were then established: first register (nothing changes), second register (desired behavior, see Figure~\ref{fig:spec}(a)) or ``other'' (higher registers, quasi-periodics, ``muffled sound", see Figure~\ref{fig:spec}(b) and \ref{fig:spec}(c)).

Some holes exhibit an unambiguous behavior.
This is the case of the two holes with the smallest diameter (U$_{1.0}$ and D$_{1.0}$). 
The opening of these holes does not destabilize the first register, but slightly diminishes the amplitude of the sound.
For the upstream holes, second register was systematically produced for diameters $2.4$, $3.0$ and $5.0$~mm.
Hole U$_{1.5}$ also produced second register 48 times over 56. 
There was one ``squeak'' of fourth register and seven first registers.
It is interesting to notice that one participant stayed on the first register three times over four.
This consistency suggests a different playing technique from the other participants. 
In a musical context, hole U$_{1.5}$ would not allow this participant to play twelfths.

For the downstream holes, tendencies are slightly more blurred.
One the one hand, when the diameter increases, the proportion of first register decreases.
For the hole D$_{2.4}$, the first register persists only two times over 56.
The proportion of ``other'' behaviors is also increasing with the diameter.
For the hole D$_{5.0}$, there is 0$~\%$ of first or second register.
Behaviors noted include third and fourth registers (Figure~\ref{fig:spec}(b)), and muffled sound (Figure~\ref{fig:spec}(c)).
This last behavior corresponds to the first register of tube D$_{5.0}$, whose first impedance peak has a much higher frequency than the closed tube's (253~Hz against 156~Hz), as illustrated on Figure \ref{fig:2}(b).

On the other hand, the emergence of second register is less clear for downstream holes. 
The proportion of second register seems to pass through an optimum for the 2.4~mm diameter. 
The real optimal diameter may lie between 2.4~mm and 3.0~mm.
Nevertheless, the high proportion of other behaviors suggests a lower stability region for the second register, or a smaller basin of attraction compared with the upstream holes.

\subsection{Numerical simulations}
\subsubsection{Simulation context}

Numerical simulations were carried out with $N=12$ modes, following the same procedure as the experiment: starting from the closed hole with constant control parameters $\gamma, \zeta, \omega_r$ and $q_r$, the hole is opened if the first register is stable.

To encompass the control parameters space as exhaustively as possible, the space $(\gamma,\zeta,\omega_r)$ is mapped by 1000 latin hypercube samples.
Simulations are repeated for $q_r\in[0.1; 1]$, in steps of 0.1.
The ranges of the control parameters are reported in Table~\ref{tab:2}.

Simulations are divided in two phases. 
In the first, the hole is closed: modal coefficients $C_n$ and $s_n$ are taken from the closed hole impedance.
Initial conditions are set to zero, except for the first modal pressure $p_1(t=0)=0.5$.
In this way, the stability of the first register is favored.
This first phase lasts 0.5~s.
At the end, the register is determined from the modal pressure of maximum amplitude, in the absence of quasi-periodics.
Only the points of the control parameters space leading to first register are kept for the second phase.

In the second phase, the hole is instantly opened: modal coefficients change to the values of $\tilde{C}_n$ and $\tilde{s}_n$ that account for the new configuration of the resonator and the nonlinear losses in the newly open hole.
Simulations are carried out with and without localized nonlinear losses.
In the first case, $\tilde{C}_n$ and $\tilde{s}_n$ vary with respect to $v_\mathrm{RMS}$, unlike the second case where they remain constant.
The initial conditions of the second phase are the final conditions of the first phase.
The second phase lasts 1.5~s.
The register is determined at its end.

\begin{table}
	\centering
	\caption{Range of the parameters used during numerical simulations.}
	\label{tab:2}
	\begin{ruledtabular}
	\begin{tabular}{lcccc}
		& $\gamma$	& $\zeta$ & $\omega_r/(2\pi)$ [Hz] & $q_r$ \\
		\hline
	Min	& 0.05 & 0.05 & 1000 & 0.1 \\
	Max & 2.2 & 0.6 & 2500 & 1
	\end{tabular}
	\end{ruledtabular}
\end{table}

\subsubsection{Cartography in the control parameters space}

\begin{figure}[h!]
	\centering
	\includegraphics[width=.4\textwidth]{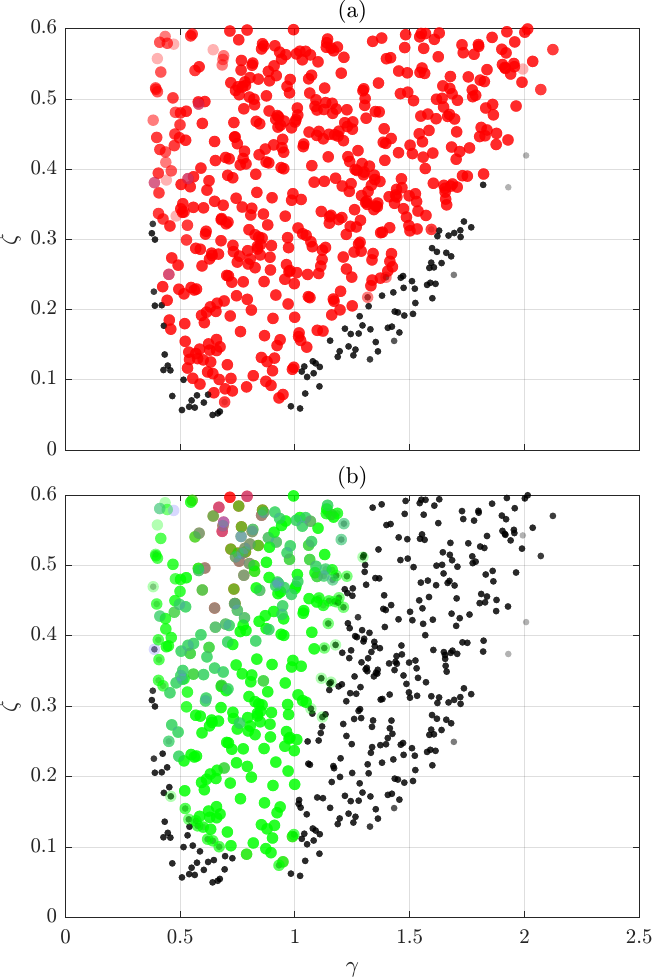}
	\caption{Cartography of the register obtained after the opening of hole U$_{2.4}$, in the 4-D control parameters space projected on the $(\gamma, \zeta)$ plane. Simulations without (a) and with (b) localized nonlinear losses in the register hole. Black, red, green and blue dots respectively refer to the equilibrium, the first and the second register, and upper registers.}
	\label{fig:carto}
\end{figure}

The time-integration scenario previously described is applied to each of the ten open holes, using the Matlab solver \texttt{ode45} with absolute and relative tolerances of $10^{-8}$.
The registers produced at the end of the simulations are represented  for the open hole U$_{2.4}$ on Figure~\ref{fig:carto}.
For readability, the points of the 4D space are projected on the $(\gamma, \zeta)$ plane.

When nonlinear losses are not considered (Figure~\ref{fig:carto}(a)), the opening of the hole does not change the stability of the first register for the most part.
Some points located on the edges of the cartography  converge towards equilibrium.
For these points, the initial conditions of the second phase are no longer located in the basin of attraction of the first register. 
Therefore, the first register is not maintained in these cases. 
However, note that no conclusion can be drawn on the stability limit of the first register.

The results are markedly different when nonlinear losses are considered (Figure~\ref{fig:carto}(b)). 
Firstly, the region of oscillating regimes is much smaller than in the previous case. 
Secondly, the first register has almost been replaced by second register. 
Some points of first register remain stable for very high values of $\zeta$.
Some of them are superimposed with second and higher registers (blue dots) when lowering the value of $q_r$.

\subsection{Comparison between the experiment and simulations}

\begin{figure}[h!]
\centering
\includegraphics[width=.45\textwidth]{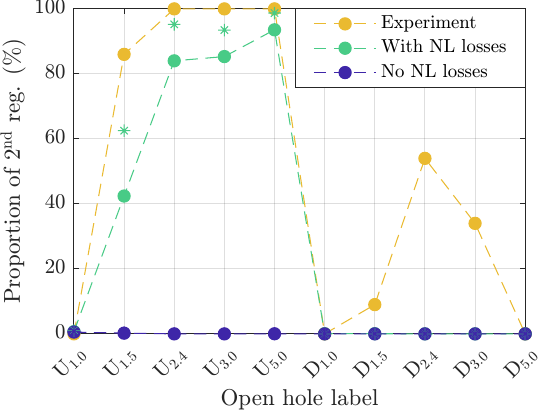}
\caption{Proportion of second register among the oscillating regimes, for each open hole. Comparison between the experiment and simulations, with and without localized nonlinear losses in the register hole. The asterisks (*) denote the proportion of second register, when considering only points of the control parameters space for which $\zeta<0.4$.}
\label{fig:propR2}
\end{figure}

Figure \ref{fig:propR2} shows the proportion of second register produced for each open hole, in the experiment and in the simulation with and without localized nonlinear losses in the register hole.

Concerning simulations without nonlinear losses,  second register was never produced, regardless of the hole considered.
In the simulations accounting for localized nonlinear losses, results are much closer to the experiment for the upstream holes.
The larger the hole, the closer the proportion of second register to 100~\%.

For hole U$_\mathrm{1.5}$, the proportion of second register remains weak. 
This can first be explained by an exceedingly wide control parameter space.
For $\zeta>0.45$, first register can be produced by the model accounting for nonlinear losses, as illustrated on Figure~\ref{fig:carto}(b).
These values are too high for a clarinet, where $\zeta$ is generally well below 0.4 \citep{fritz2004clarinette,dalmont_oscillation_2007}.
When considering only points of the parameter space for which $\zeta<0.4$, the proportion of second register jumps of 20~\% for hole U$_\mathrm{1.5}$ and becomes closer to the experimental results for the other upper holes, as shown by the asterisks on Figure \ref{fig:propR2}.
The same would apply for $q_r=0.1$ and the production of upper registers, which corresponds better to a reed organ pipe \citep{silva_interaction_2008} than to a clarinet.
When restraining to $q_r\geq 0.2$, the proportion of second register reaches nearly 100~\% for holes U$_\mathrm{2.4}$, U$_\mathrm{3.0}$ and U$_\mathrm{5.0}$, which brings them much closer to the quantitative experimental results.

For the downstream holes, none of the two simulation methods succeeded in producing second register.
The following discussion proposes explanations for the differences in second register production between the experiment and the model accounting for nonlinear losses.

\subsection{Discussion}
The model accounting for nonlinear losses exhibits promising results for the upstream holes. 
For the downstream holes, where the experiment has shown a wide variety of behaviors, the model apparently fails to produce second register. 
This raises the question of the stability of the second register for the model. 
Therefore, to test the stability of the second register, simulations were carried out, but this time imposing the condition $p_2=0.5$ and $p_n=0$ for all other modal pressures when opening the hole.
The points in the $(\gamma, \zeta, \omega_r)$ space were the same as in the previous simulations, with $q_r=0.4$. 

\begin{table}[h!]
	\centering
	\caption{Percentage of second register obtained by imposing an initial condition promoting its production when the hole is opened, for both models. Simulations for $q_r=0.4$ and $\zeta<0.4$.}
	\label{tab:3}
	\begin{ruledtabular}
	\begin{tabular}{lcc}
		Hole label & With NL losses & Without NL losses\\
		\hline
		D$_{1.5}$ &  73.6~\% & 68.3~\%\\
		D$_{2.4}$ & 57.0~\% & 39.0~\%\\
		D$_{3.0}$ & 38.2~\% & 28.6~\% \\
		D$_{5.0}$ & 0.0~\% & 0.0~\% \\
	\end{tabular}
	\end{ruledtabular}
\end{table}

The results from these simulations are presented in Table~\ref{tab:3}, showing that stable second register can exist for holes D$_\mathrm{1.5}$, D$_\mathrm{2.4}$ and D$_\mathrm{3.0}$.
For the points where stable second register is found, the $1^\mathrm{st}$ and second  registers are therefore multistable.
Moreover, the second register stability region is wider for the model accounting for localized nonlinear losses than without.  

In the multistability region, the emergence of the second register is conditioned in the phase space by its basin of attraction, i.e. the set of initial conditions leading to the second register.
Thus, for the points of the cartography that do not lead to the second register, initial conditions are located in the phase space outside the basin of attraction of the second register.
In the experiment, second register was sometimes produced for downstream holes.
By analogy with the model, when the hole was opened, the initial conditions were located in the basin of attraction of the second register.

Some contributions to the simulation could help stimulate the production of second register.
First, a note transition model such as the one from \citet{guillemain2005dynamic} or \citet[Chap. 6.6.2]{taillard2018phd} could enable the system to reach the inside of the second register's basin of attraction.
Furthermore, choosing different final conditions along the limit cycle of the first register during the first phase (closed hole) would provide the second simulation phase (open hole) with a set of initial conditions more evenly distributed in the phase space.
Some of them could fall into the second register's basin of attraction.

\section{Conclusion}

This study shows that in the context of sound synthesis of clarinet-like instruments by modal decomposition of the input impedance, it is necessary to account for localized nonlinear losses in the register hole during note transitions leading to second-register notes.

An experiment with a clarinetist establishing a first register regime and an operator opening a hole along the tube highlights the geometrical characteristics of an ``efficient" register hole, i.e. one that provokes the transition to second register regime when opened.
Results show that when opening a hole located upstream, participants systematically produced second register from a sufficiently large diameter.
For downstream holes, other registers are observed, particularly for large-diameter holes.
The proportion of second register passes through an optimum for an intermediate hole diameter.

The experiment is replicated numerically, by time integration of a clarinet-type system of equations.
Two different models are tested.
The first is the standard model based on the modal decomposition of the input impedance of the resonator.
The second accounts for localized nonlinear losses in the register hole, through the nonlinear losses model for side holes from \citet{dalmont_experimental_2002}.
These losses are integrated into the physical model using a variable modal coefficients method \citep{szwarcberg2023amplitude,diab2022nonlinear}.
In the case of the model without nonlinear losses, simulations never produce second register, for any of the open holes.
Concerning the model with nonlinear losses, the proportion of second-register production is close to the experiment for upstream holes, but remains at zero for downstream holes.

In the context of sound synthesis by modal decomposition of the input impedance, the proposed method could enable to better transcribe the dynamical behavior of second-register notes.
This possibility requires similar studies on the clarinet's entire second register, i.e. from B$_4$\textsuperscript{\ref{note:1}} to C$_6$.
For each of these notes, the relative longitudinal position of the register hole shifts with respect to its ideal position, located at the node of the second mode shape.
This shift affects the ease of production of the second-register regime in a manner that could be quantified experimentally, and precisely described in a simplified control space by a numerical model.

Finally, this study opens up new perspectives on understanding the phenomena behind the loss of stability of the first register when the register hole is opened.
Future studies could investigate the impact of localized nonlinear losses on the multistability of the oscillating registers, as well as the evolution of the size of their basins of attraction.

\section*{Acknowledgments, Author declarations and Data availability statement}
This study has been supported by the French ANR LabCom LIAMFI (ANR-16-LCV2-007-01). 
The authors warmly thank \'E. Gourc for the drilling of the holes, as well as all the participants to the experiment.
The authors have no conflicts to disclose.
The data that support the findings of the present study are available from the corresponding author upon reasonable request.
The authors declare that the experiment was carried out in conformity with the Ethical Principles of the ASA.

\bibliography{biblio}
\newpage 
\listoffigures
\listoftables
\end{document}